\documentclass[acmsmall]{acmart}

\copyrightyear{2022}
\acmYear{2022}
\setcopyright{rightsretained}

\acmJournal{PACMHCI}
\acmYear{2022} \acmVolume{6} \acmNumber{CSCW2} \acmArticle{395} \acmMonth{11} \acmPrice{} \acmDOI{10.1145/3555120}

\usepackage{etoolbox}
\makeatletter
\patchcmd{\maketitle}{\@copyrightpermission}{
   \begin{minipage}{0.2\columnwidth}
    \includegraphics[width=0.90\textwidth]{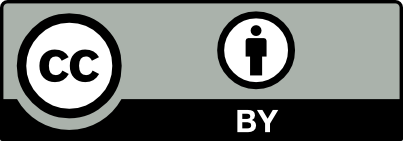}
   \end{minipage}\hfill
   \begin{minipage}{0.8\columnwidth}
     This work is licensed under a Creative Commons Attribution International 4.0 License.
   \end{minipage}

   \vspace{5pt}
}{}{}

\makeatother

\usepackage{pifont}
\newcommand{\cmark}{\ding{51}}%
\newcommand{\xmark}{\ding{55}}%
\usepackage{multirow}
\usepackage{siunitx}
\usepackage{enumitem}
\definecolor[named]{mygreen}{cmyk}{0.53,0,0.58,0.43}
\definecolor[named]{myred}{cmyk}{0.21,0.63,0.38,0}
\def\BarGreen#1{
  {\color{mygreen!100}\rule{#1mm}{8pt}}}
\def\BarRed#1{
  {\color{myred!100}\rule{#1mm}{8pt}}}

\begin{document}

\title[The Gift That Keeps on Giving]{The Gift That Keeps on Giving: Generosity is Contagious in Multiplayer Online Games}

\author{Alexander J. Bisberg}
\authornote{Both authors contributed equally to this research. \newline
This work is partially supported by NSF (award no. 2035064) and DARPA (award no. HR001121C0169).}
\email{bisberg@usc.edu}
\author{Julie Jiang}
\authornotemark[1]
\email{yioujian@usc.edu}
\affiliation{%
  \institution{University of Southern California}
  \country{United States}
 }

\author{Yilei Zeng}
\affiliation{%
  \institution{University of Southern California}
  \country{United States}
}
\email{ yileizen@usc.edu }

\author{Emily Chen}
\affiliation{%
  \institution{University of Southern California}
  \country{United States}
 }
\email{echen920@usc.edu}

\author{Emilio Ferrara}
\affiliation{%
  \institution{University of Southern California}
  \country{United States}
 }
\email{emiliofe@usc.edu}

\renewcommand{\shortauthors}{Alexander J. Bisberg et al.}

\begin{abstract}
Understanding social interactions and generous behaviors have long been of considerable interest in the social sciences community.
While the contagion of generosity is documented in the real world, less is known about such phenomenon in virtual worlds and whether it has an actionable impact on user behavior and retention. 
In this work, we analyze social dynamics in the virtual world of the popular massively multiplayer online role-playing game (MMORPG) \textit{Sky: Children of Light}. 
We develop a framework to reveal the patterns of generosity in such social environments and provide empirical evidence of social contagion and contagious generosity.
Players become more engaged in the game after playing with others and especially with friends.
We also find that players who experience generosity first-hand or even observe other players conduct generous acts become more generous themselves in the future.
Additionally, we show that both receiving and observing generosity lead to higher future engagement in the game.
Since \textit{Sky} resembles the real world from a social play aspect, the implications of our findings also go beyond this virtual world.
\end{abstract}


\maketitle

\section{Introduction}
Human beings are a naturally social species that hinges on cooperation and collaboration for survival \cite{nature2018cooperative}. 
From time to time, however, we may also seek companionship simply for our mental and emotional well-being \cite{coan2015social}. 
Socialization plays an integral part in not only offline relationships, but also in online relationships through mediums such as social media \cite{garton1997studying} and multiplayer online games \cite{yee2006psychology,Cole2007,Zhong2011}: these studies show that people develop personal, close, and even intimate relationships with others online, even if they have never actually met in real life. 
In fact, the urge to socialize is what drives some people to participate in online activities in the first place \cite{yee2006motivations,whiting2013people,hassouneh2014motivation}.

An important aspect of social interaction is generosity, or acts of kindness. 
There are many documented instances in which people display altruism for others without the expectation of being reciprocated or socially rewarded \cite{tsvetkova2014social,batson1978altruism,post2005altruism}. 
Researchers explain these behaviors with frameworks of evolutionary biology, physiological models, and positive psychology \cite{post2005altruism,sommerfeld2010subjective}. 
Witnessing or experiencing acts of kindness can also propel a person to be generous towards others (pay-it-forward), a form of social contagion of behavior \cite{pressman2015pif, jung2020prosocial,tsvetkova2014social,Kizilcec2018social,Woo2015,kang2014altruism}. 

\begin{figure}
    \centering
    \includegraphics[width=0.9\linewidth]{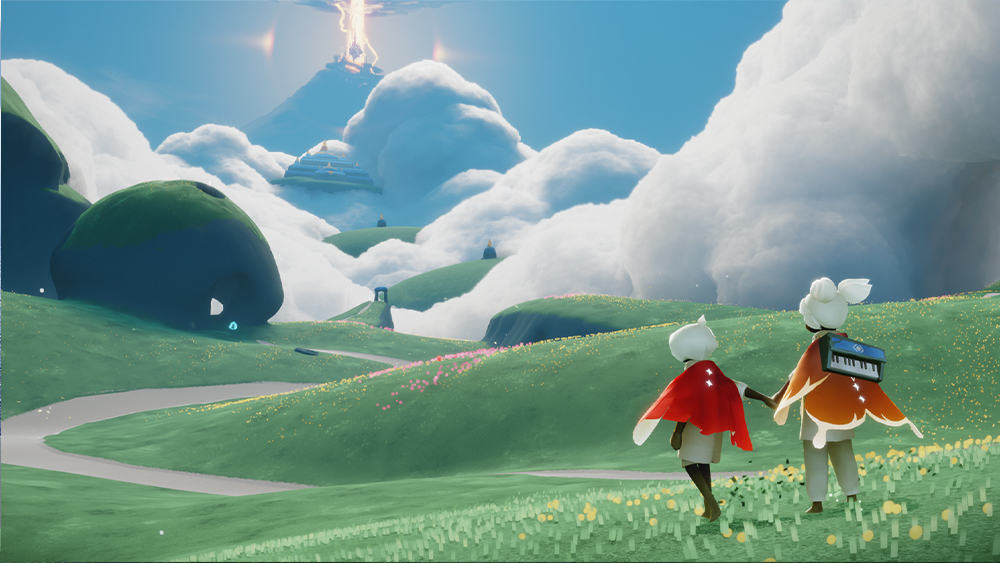}
    \caption{A stylized depiction of \textit{Sky}: Children of Light \cite{sky}.}
    \label{fig:sky}
\end{figure}

Motivated by existing literature, we conduct a large-scale data-driven empirical study of contagious generosity of nearly 1 million players on \textit{\textit{Sky}: Children of Light} \cite{sky} (Figure \ref{fig:sky}), an award-winning free-to-play open world social adventure massively multiplayer online role-playing game (MMORPG) for iOS and Android \cite{apple2019best}. 
In \textit{Sky}, players interact with each other in an expansive virtual world. However, exploration is only one part of the game. 
Players are matched together in this virtual world, unlocking new areas through collaborations and relationship-based progression systems. 
Beyond chatting and social play, a central aspect of \textit{Sky} lies in its mechanism to facilitate gift/currency exchange between players using candles to unlock friendship features.

Our framework for revealing patterns of generosity is distinct from previous work on generosity for a number of reasons.
First, \textit{Sky} is an online game with millions of players \cite{gamasutra2020}, allowing us to expand the sample size far beyond similar studies \cite{tsvetkova2014social}.
Second, \textit{Sky} is a unique online platform in that the generous behavior of gifting candles occurs in public spaces and is, by design, not necessarily reciprocal.
This enables us to study both generalized reciprocity (directly benefiting from generosity) and third-party influence (witnessing generosity happening between other people). 
Third, players in \textit{Sky} are not measured by a specific success metric: one cannot be \textit{better} at this game than another, only \textit{relationships} can be leveled up. 
Therefore, we argue we are observing social play and pure acts of kindness in an environment free of ulterior motives or added incentives. 
Finally, while studies have shown that social dynamics impact communication and collaboration in the virtual world \cite{ducheneaut2004social,ducheneaut2007virtual,nardi2006strangers,Cole2007}, to the best of our knowledge, our work is the first to demonstrate that social contagion of generosity can impact user progression and retention.

The main findings of this paper are as follows:
\begin{enumerate}

\item Social play with friends is a \textbf{positive predictor} of future engagement, while solo play is a negative predictor.
\item Players benefiting from or observing generous acts are likely to \textbf{engage more} with the game in the future.
\item Players benefiting from or observing generous acts are likely to be \textbf{more generous} towards others.

\end{enumerate}

In this paper, we begin by grounding our motivations and theoretical framework through surveying related work and introducing our research questions (\S\ref{sec:related_work}), followed by a description of the game and the data we used (\S\ref{sec:sky}). 
We then structure the analytical portion of the paper into two parts: the first involves understanding how social contagion impacts future engagement (\S\ref{sec:social_contagion}) and the second involves our quasi-experimental approaches to study whether generosity is contagious and how that impacts the recipient's future generosity as well as future engagement (\S\ref{sec:generosity}). 
Finally, we conclude our paper with some discussions and implications of our findings for the wider CSCW community (\S\ref{sec:discussion}).

\section{Related Work}\label{sec:related_work}

The majority of literature related to generosity in virtual worlds focuses on massively multiplayer online (MMO) games, or more specifically MMO role-playing games (MMORPGs), a genre in which an individual player is spawned in a virtual world to complete quests, level up their characters, and interact with other human and non-human avatars. 
In another type of MMO, multiplayer online battle arenas (MOBA), players are matched to teams and play out a battle in a small arena rather than an open world to achieve some goal, such as destroying the opponents' base. 

\subsection{Motivations for MMO Gameplay}
Existing research suggests that players' motivation for gameplay could be linked to their proclivity for social interactions in addition to other gameplay artifacts.
\citet{yee2006motivations} examined players' motivations for playing MMORPGs and proposed three categories of players: (1) achievement-oriented players who desire to advance in the game, improve their mechanics of gameplay, and compete with others, (2) social-oriented players who are drawn to socializing, meaningful relationships, and teamwork, and (3) immersion-oriented players who are interested in game discovery, role-playing, and customization. 
This work was extended by \citet{kahn2015trojan}, who refined the player's motivation into six categories: socializer, story-driven, escapist, completionist, smarty-pant, and competitor.

\subsection{Player Behavioral Modeling in MMOs}
Player behavioral modeling is a research area of interest to industry game developers and academic game researchers \cite{bakkes2012player,yannakakis2013player}. 
This technique uses computational methods to model players in games, with the aim of understanding, characterizing, and predicting player behaviors from cognitive, affective, and behavioral aspects \cite{yannakakis2013player}. 
With the availability of large-scale data collected from MMOs, empirical game analytics based on player modeling have become widespread \cite{Suznjevic2011,de2016player,Harrison2015,hooshyar2018data}. 
For example, studies on \textit{World of Warcraft} (WoW) \cite{wow}, one of the most popular MMORPGs, have shown successes in modeling player actions (trading, questing, dungeons, raiding, etc.) to predict a player's in-game engagement \cite{Suznjevic2011} or progression \cite{suznjevic2010mmorpg}. 
Later, \citet{Harrison2015} compared various techniques to predict player achievements in \textit{WoW}. 
Other works utilize player modeling for churn prediction in \textit{Tomb Raider} \cite{mahlmann2010tombraider} and  \textit{Destiny} \cite{tamassia2016destiny} or game outcome prediction in \emph{League of Legends} (\textit{LoL}) \cite{lol, jiang2020lol, sapienza2018individual} and \emph{Defense Against the Ancients 2} (\textit{DOTA2}) \cite{dota2, lan2018player}. 
Some research also utilizes data-driven methods to profile players in order to cluster and identify major categories of players in \textit{Tomb Raider} \cite{drachen2009tombraider} and \textit{LoL} \cite{sapienza2018ntf,ong2015player,jiang2021wide}.

\subsection{Player Interaction in MMOs}
In MMOs, players can fully immerse themselves in a virtual world and interact with other players through their avatars \cite{taylor2009play}. 
Many studies have shown that players form meaningful interpersonal relationships in such games \cite{yee2006psychology,Cole2007,taylor2009play,Zhong2011}. 
Moreover, quality online relationships are an important contributing factor for continued play \cite{voiskounsky2004playing,lee2010what} and enjoyment in games \cite{sweetser2005gameflow}. Positive social interactions in MMORPGs can facilitate game enjoyment, which in turn increases game engagement \cite{chen2006enjoyment}. 
At times, these online relationships can even rival offline relationships \cite{yee2006psychology}. 
Additionally, \citet{Zhong2011} found that collective play can positively influence bridging (connections \textit{between} communities) and bonding (connections \textit{within} a community) social capital, as well as online civic engagement. 

There are several studies that qualitatively investigate social interactions in online games.
\citet{ducheneaut2004social} and \citet{ducheneaut2007virtual} used the MMO \textit{Star Wars Galaxies} \cite{swg}, a game in which player interdependence and cooperation is crucial for personal progression, to examine mediums of social interaction and compare them to real-world scenarios.
 \citet{nardi2006strangers} provided an overview of collaboration types in \textit{WoW}.

A subsequent question one may ask is why in-game social interactions are so prominent in virtual worlds.
\citet{Cole2007} suggests that the anonymity provided by the online medium allows a safe environment for players to open up to other players and become emotionally involved. 
Players can feel ``more themselves'' in their online presence because they will not be judged by their physical appearance \cite{Cole2007}. 
Although our work here focuses on an MMORPG, we believe there are some interesting studies on MOBAs that are worth mentioning that specifically focus on network formation and teamwork. 
\citet{sapienza2019deep} learned co-play networks in \textit{DOTA2} and \citet{ong2015player} modeled individual playstyles in \textit{LoL}. 

All of the studies above represent how social dynamics are studied for the sake of understanding communication and collaboration in virtual worlds; however, there remains a gap in the literature in using these interactions to predict progression and exploration.
We provide explanations and insights into how in-game social interactions contribute to future player engagement. 
Further, player-to-player interaction is not strictly necessary in \textit{Sky}, which makes it an appropriate medium to study authentic and genuine social relationships motivated by the intrinsic desire to connect with others. 
Existing research suggests that social behavior in the virtual world has a significant impact on social behavior in the real world \cite{qin2011influence}.
Since social relationships in the virtual world of \textit{Sky} mimic real-world relationships, our findings have bigger implications in understanding engagement in real-world situations.

\subsection{Social Contagion in Virtual Worlds} \label{sec:soc_cont}

Social contagion is the phenomenon that emotions, beliefs, and behaviors can spread through social networks much like outbreaks of diseases \cite{marsden1998memetics}. 
Some studies have shown how certain negative actions in games are socially contagious.
\citet{woo2013contagion} provided evidence of the contagion of cheating by using bots in the game. 
\citet{shen2020viral} showed that toxic behaviors are also contagious.

Another major research area focuses on how social contagion is linked to monetization and player engagement in games. 
\citet{zeng2019influence} showed that teammates can exert peer influence on one another in team-based games\cite{zeng2019influence}.  
In-game purchases can also be directly or indirectly peer-influenced \cite{geng2015social, zeng2020learning}, with stronger social ties having more impact than weaker social ties \cite{zhao2017social}.
\citet{fang2019social} discovered that while \emph{pure friends} (friends with no common friends) and \emph{Simmelian-tie friends} (friends of friends) can both peer influence a player to spend money, pure friends have a stronger impact. 
Research on the effect of social contagion on player engagement showed that influential players have an outsized impact on the playtime---in particular, social play---of other players \cite{ki2014identifying,canossa2019influencers}, indicating that influential players are socially contagious and important players. 

In this work, we operationalize social contagion as contagious social play with friends or strangers. 
We pose the following research question about social contagion in \textit{Sky}:
\begin{enumerate}[labelindent=\parindent,leftmargin=*,nosep,label=\textbf{RQ\arabic*:}]
    \item Does social contagion impact future engagement?
\end{enumerate}

\subsection{Contagious Generosity in Virtual Worlds} \label{sec:cont_gen}

It is well known that generosity is a contagious human behavior \cite{tsvetkova2014social,pressman2015pif}, meaning those who receive or observe a generous act are more likely to pay it forward by being generous to others. 
There are two distinct types of generosity contagion \cite{tsvetkova2014social}:
\begin{itemize}
    \item \textit{Generalized reciprocity}: the recipient of the generous act is likely to become more generous.
    \item \textit{Third-party influence}: an observer of a generous act will become more generous.
\end{itemize}
Such effects have also been observed in virtual environments.
Research on generosity in the MMORPG \textit{Aion} \cite{aion} revealed two types of generous acts in such games: gifting in-game assets and shepherding lower-level players. 
Players on the receiving end of these acts of kindness are observed to have an increase in their generosity as well as in their loyalty and satisfaction with the game \cite{Woo2015,kang2014altruism}. 
In another virtual setting, \citet{Kizilcec2018social} found that receiving virtual gifts on \textit{Facebook} made the recipient more likely to pay it forward. 

In this paper, we test both theories of generalized reciprocity and third-party influence by examining the effects of gift-giving in \textit{Sky}, which is a prominent part of the game experience. 
As such, we raise the following two research questions:

\begin{enumerate}[labelindent=\parindent,leftmargin=*,nosep,label=\textbf{RQ\arabic*:}]
    \setcounter{enumi}{1}
    \item How does receiving and/or observing generosity impact a player's future generosity?
    \item How does receiving and/or observing generosity impact future engagement?
\end{enumerate}

In particular, we consider both the players' generosity in the future, as measured by how likely they will offer gifts to someone else (pay it forward), as well as future engagement, as measured by the duration of time they spend in the game engaging in various activities.
We discuss the exact mechanisms of in-game progression and gift-giving in \S\ref{sec:sky}.

\section{Sky: Children of the Light} \label{sec:sky}

\subsection{Background}
\textit{Sky} \cite{sky} was released in July 2019 by \textit{ThatGameCompany} (TGC) and topped 50 million downloads worldwide  by 2020 \cite{gamasutra2020}. 
Players in \textit{Sky} enjoy an immersive exploration of a virtual world, balancing their playtime in both goal-oriented and open-ended play \cite{kerr2020using}.
The game brings players together to share intimate moments by artificially creating boredom in the game and removing the emphasis on goal-oriented play;  
in slowing the gameplay, the game pushes players to engage creatively and socially with other players \cite{kerr2020using}. We describe several key aspects of the game below.

\paragraph{Exploration.} 
Players explore the world of \textit{Sky} through an avatar on foot, interacting with `spirits' to unlock emotes such as waving, farming wax to create candles, and progressing through levels. 
Candles are a limited resource in the game that is required to unlock social emotes, cosmetics, and other items.
Players can also fly in the world using their capes. 
There are permanent power-ups called ``wing buffs'' that grant longer fly times.

\paragraph{Friendship.} When players enter the virtual world of \textit{Sky}, they are matched with other players.
When two players meet for the first time, they appear as grayed-out silhouettes to each other. 
Their true avatars are only unveiled when one tries to establish a formal friendship through candle exchange: one player must offer a white candle (generated from farmed wax) and the other must accept, a process known as `exchanging light'.
If the candle is not accepted by the other player, they will remain silhouettes to each other.
There are a variety of interactions players can engage in with friends, such as chatting, giving high-fives, or hugging. 
One of the signature experiences of \textit{Sky} is to hold hands with friends, which allows one friend to guide the other as they explore the world together. 
Beyond pure virtual friendships developed in \textit{Sky}, players can also invite real-life friends to play the game.

\paragraph{Social Benefits}
Playing with friends has many important benefits in \textit{Sky}.
These include hastening energy recharge time, extending flight, and  enabling the opening of multiplayer doors to access secret levels.
Players can also warp directly to the location of their friends when they are in the same session of gameplay.

\paragraph{Anonymity.} \textit{Sky} takes measures to preserve anonymity among players. 
Besides the aforementioned greyed-out silhouettes seen by players who have not established a friendship, many other aspects of player identity are anonymized. 
Players do not explicitly choose a display name for other players to see, nor can they see other players' names by default.
When two players become friends each of them has the opportunity to give the other a nickname or let the game randomly choose one. 
Moreover, until two players are friends, chat messages appear as ellipses and cannot be read unless they use a special chat prop like a picnic table or swing bench.
These can be found in the environment or deployed by players. 
The baseline communication method is emotes performed by player avatars.

\paragraph{Customization.} 
Players are able to customize their in-game avatars in a variety of ways. 
For instance, players can find and collect emotes across the realm, which can be used to express themselves (e.g., waving, shivering). 
Similarly, they can also collect cosmetics to further customize their outfit, hairstyle, instruments, etc.

\subsection{Types of Play}
Playstyles in \textit{Sky} are jointly determined by the experience type and the activity type. The granularity of playstyles is determined on a session-level basis. \textit{Sky} defines a session as any continuous period of gameplay with less than 60 seconds of break time in between, and session-levels are disjoint parts of a session separated by distinct virtual spaces of the game.
In \textit{Sky}, there are three categories of experience types for every player in every session-level:

\begin{itemize}
    \item \textbf{Solo}: No other players are considered nearby in the game, where the definition of nearby is when two players are at most 1 unit (the distance of a player avatar) away from each other. 
    \item \textbf{Group (stranger)}: At least one other player is nearby, but none of them are friends.
    \item \textbf{Group (friend}): At least one other player is nearby and one of them is a friend.
\end{itemize}

The distinction between a group (friend) and a group (stranger) session is subtle but important. In particular, a group (stranger) session can become a group (friend) session in one of two ways: the players can undergo the candle exchange process to become friends, or they can teleport directly to a friend's location in \textit{Sky}.

In each session-level, players are able to engage in  \textbf{social}, \textbf{progression} activities, both, or neither. 
Social session-levels are session-levels in which players engage in social interactions with other players and usually involve candle-giving.
By definition, they can only occur in group (vs. solo) session-levels. 
Progression session-levels are session-levels in which players complete activities such as quests or wax farming.
When a player is neither socializing nor progressing in a session-level, their activity is deemed to be idle/unknown.
This acts more as a catch-all as opposed to a well-defined gameplay state.
In combining experience types and activity types, we are able to characterize a player's playstyle for that session-level.

\subsection{Data} \label{sec:emp_data}

\begin{figure}
    \centering
    \includegraphics[width=0.9\linewidth]{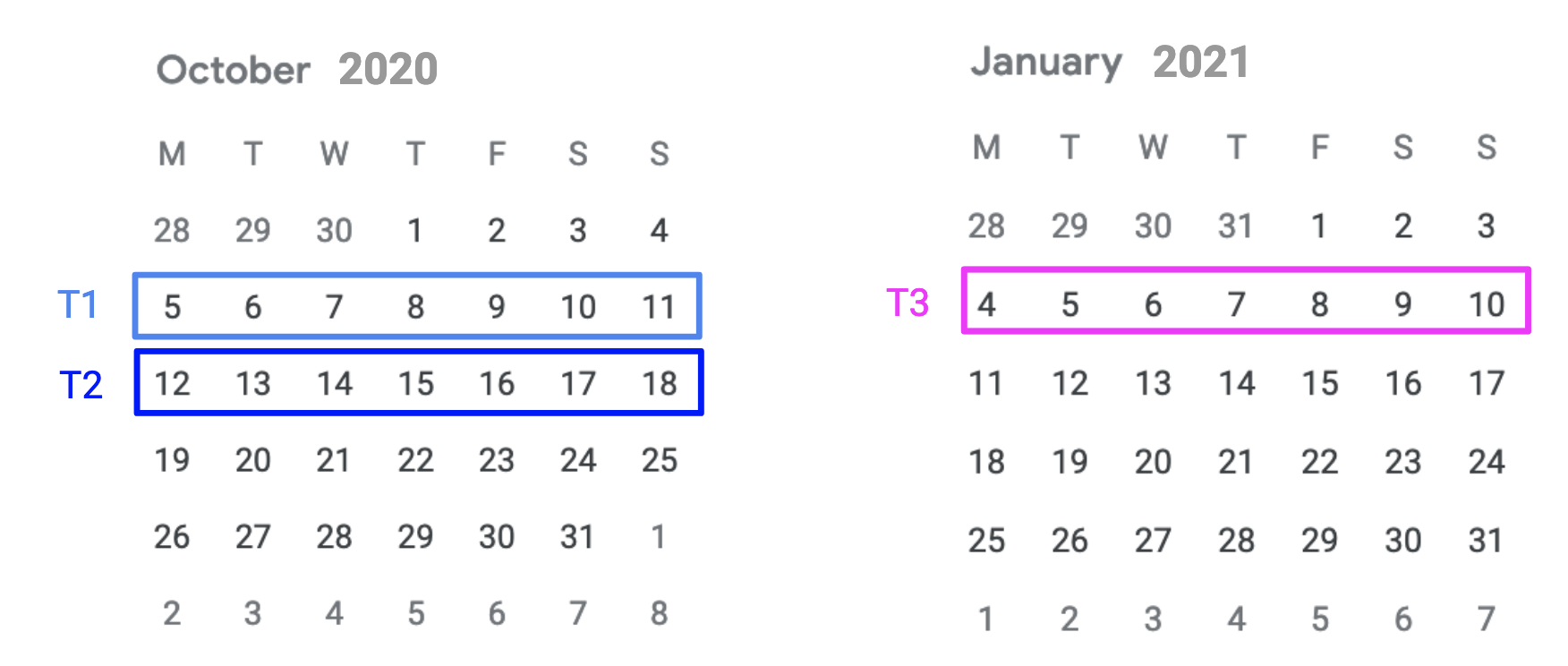}
    \caption{Date ranges for the experiments we performed. 
    T1: Monday, October 5, 2020, to Sunday, October 11, 2020.
    T2: Monday, October 12, 2020, to Sunday, October 18, 2020.
    T3: Monday, January 4, 2020, to Sunday, January 10, 2020.}
    \label{fig:times}
\end{figure}

Due to a collaboration with TGC, we have access to detailed player and session-level data.
We collected data from Monday, October 5, 2020, to Sunday, October 11, 2020 (T1 - see Figure \ref{fig:times}).
We deliberately chose a time period that avoids most major holidays, as people tend to be more cheerful \cite{nyt2012happy} and generous \cite{wp2018generous} than normal. 
This time period allows us to observe generosity among players without the influence of external stimuli.
Our data collection from this time frame yields 939,000 players and 11 million sessions. 
We remove 45,000 players who are deemed inauthentic internally by \textit{Sky} due to suspicion of cheating. 
We further exclude players who were not ``active'' during this time period, which we define as players who either had a total number of sessions or total playtime that is in the bottom 10th percentile of all players during this time frame. 
This enables us to focus our analysis on players who were actively playing \textit{Sky} during this time frame and mitigate any biases from players who quit shortly after their first session  or who were novice players adapting to the game. 
The bottom 10th percentile cutoff of the number of sessions is 1 and the total playtime is 84 minutes. 
After filtering for inactive players, our dataset contains 602,000 players and 9.5 million sessions. 
We also collect full gameplay data for these 602,000 players in the following week (October 12-18, 2020 - T2) and a week in three months (January 4-10, 2021 - T3) to perform short-term and long-term retention analysis.
See Figure \ref{fig:times} for a graphical representation of the date ranges.

\section{The effects of social contagion}\label{sec:social_contagion}

In this section, we explore the effects of social contagion on players' engagement to answer \textbf{RQ1}. 
We first consider how social contagion influences the immediate session that follows, and then examine its effects on short-term and long-term engagement. 

\subsection{Impact of Social Sessions on the Next Session}

The goal of this experiment is to determine whether there is a significant difference in immediate gameplay behavior for social sessions versus solitary sessions. 
As mentioned previously, players in \textit{Sky} are matched with strangers as they explore the world. 
The virtual spaces in \textit{Sky} are large enough so that players have the choice to not interact with others; however, they cannot control whether other players approach them.

This analysis was performed on the T1 time-frame of data explained in \S\ref{sec:emp_data}.
A session of gameplay is made up of multiple levels of distinct virtual spaces that a player traverses.
We aggregated the data on a session granularity and characterized it based on the social interactions within that session through all levels.
We implemented the following division scheme for social levels in \textit{\textit{Sky}}: 
\begin{enumerate}
    \item \cmark\ \textbf{Friend level} - Had at least one friend level.
    \item \xmark \ \textbf{Friend level} - Had no friend levels.
    \item \cmark \ \textbf{Stranger level} - Had at least one stranger level.
    \item \xmark \ \textbf{Stranger level} - Had no stranger levels.
\end{enumerate}

Going by our definition, (3) and (4) are a subset of (2) since not having a friend level could include having stranger levels. 
Across these four partitions of players, we test to see if there were significant differences in the following gameplay metrics:

\begin{itemize}
    \item \textit{Session duration}: the amount of time spent playing in minutes.
    \item \textit{Number of levels}: the number of different levels the player went to in a session.
    \item \textit{Wax farmed}: the amount of wax that is collected. Wax is the ingredient used to forge candles, which can unlock many social features.
    \item \textit{New friends}: the number of new friends made in the session, which involves a mutual exchange of candles.
\end{itemize}

To analyze how social interactions influence immediate future behavior in the game, we first measure the social composition of a player's first session in the collected dataset and then observe what happened in the player's \emph{subsequent} session. 
The experimental results, along with player population sizes, are listed in Table \ref{tab:social_results}.
Since these distributions are heavily skewed to the right, we use two-sided Mann-Whitney $U$ tests to determine differences in the distributions. 
Our results reveal that players who had more social experiences engaged significantly more in terms of session duration, number of levels, and wax farmed.

\begin{table}[]
    \centering
    \begin{tabular}{lllll}
        \toprule
        & \multicolumn{2}{c}{\textbf{Friend level}} & \multicolumn{2}{c}{\textbf{Stranger level}}\\
        \cmidrule(lr){2-3} \cmidrule(lr){4-5}
         {} & \multicolumn{1}{c}{\cmark} &  \multicolumn{1}{c}{\xmark} & 
              \multicolumn{1}{c}{\cmark} &  \multicolumn{1}{c}{\xmark}\\
         \midrule
         $N$ & $243,611$ & $342,269$ & $237,918$ & $104,351$ \\ 
        Session duration (min) & 33.5*** & 20.8*** & 22.4*** & 17.2***  \\
        Number of levels       & 7.45*** & 4.70*** & 5.19*** & 3.59*** \\
        Wax Farmed             & 209*** & 104*** & 113*** & 78.3***  \\
        New Friends            & 0.122*** & 0.140*** & 0.120*** & 0.186*** \\
         \bottomrule
    \end{tabular}
    \caption{Mean values for variables tested in players' subsequent sessions. ***$p<0.001$}
    \label{tab:social_results}
\end{table}

The subsequent number of levels visited and total playtime increase with prior social engagement, which supports the existing literature \cite{suznjevic2010mmorpg}. 
In addition, prior social engagement increases the non-social activity of wax farming.
Wax farming in \textit{Sky} is a solitary venture, as one cannot collaboratively farm wax. 
We find that those who participated in at least one group (stranger) session farmed significantly more wax in their next session compared to those who only had solo play.
One plausible theory is that the players who witnessed other players collecting wax were inspired to also collect for themselves. 
Some players who were unfamiliar with the game can also learn through observation.
For instance, players may not realize that they can farm wax to make candles until they see someone else do it, which can lead to them performing similar actions in the future.

The number of new friends made in the next session was significantly lower if players had played socially, either with friends or strangers.
An explanation for this could be that, when playing socially, players may be more inclined to join their existing friends again for the next session, which could lead them to reach out to fewer strangers. 

\subsection{Predicting Future Engagement From Playstyles}
\begin{figure}
    \includegraphics[width=0.9\linewidth]{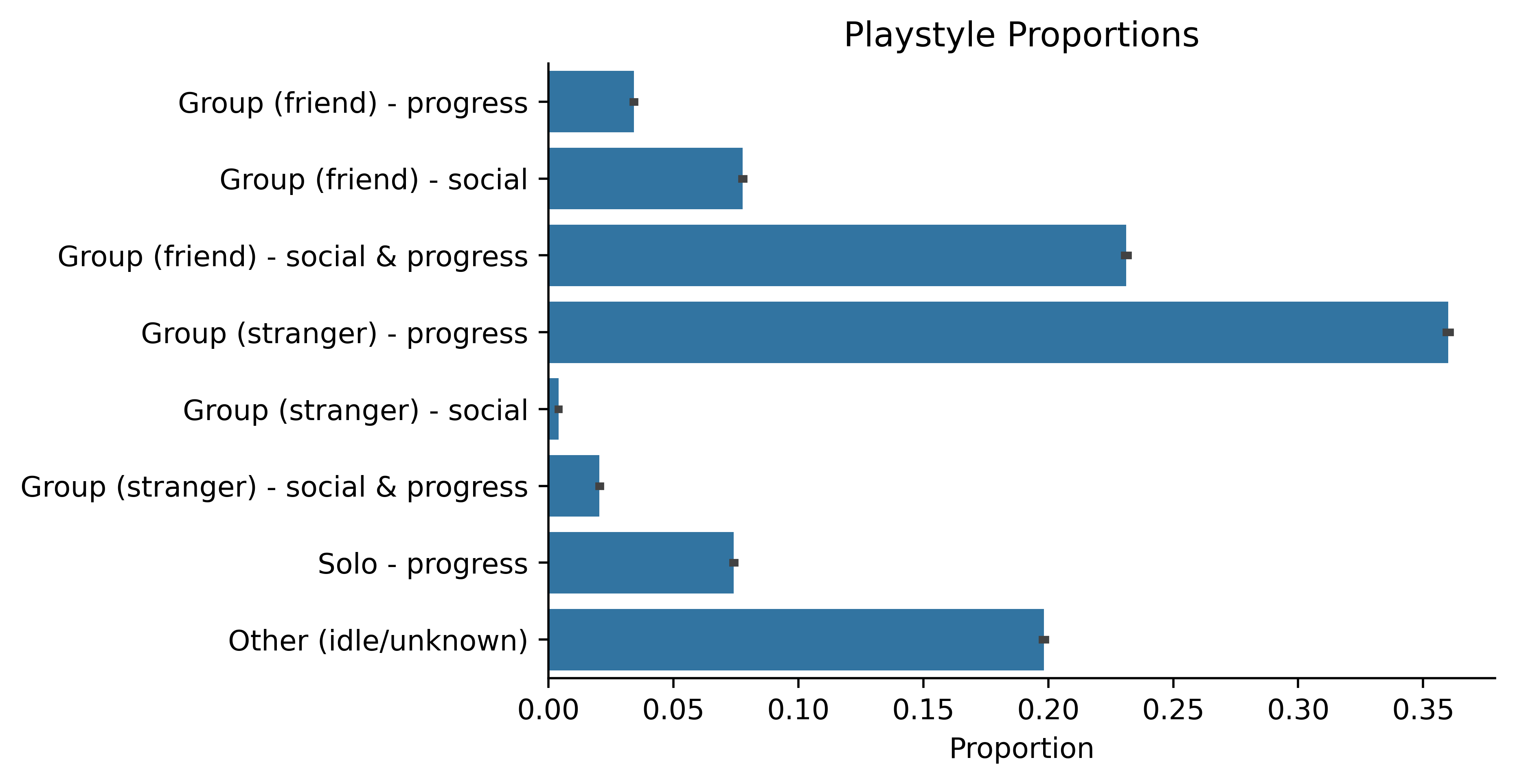}
    \caption{The average proportion of play time spent in each category of play type in T1}
    \label{fig:playstyles}
\end{figure}

We now consider the effects of a player's playstyles on their future retention and engagement in the short and long term. 
To capture a player's playstyle, we calculate the amount of time they spend in each category of play.
Figure \ref{fig:playstyles} illustrates the proportion of playtime allocated to each type of gameplay for each person on average. 
The most popular playstyle is progress sessions with strangers, which accounts for more than 25\% of total gameplay time, followed by social and progress sessions with friends (20\%). There is a clear distinction between what players tend to do with strangers versus friends. 
When playing with strangers, players tend to have progression activities and very little pure social time. With friends, however, players engage in more social activities.

To explore how playstyles affect future engagement, we use playstyle proportions in T1 as independent variables to predict gaming behavior in two target weeks: T2 and T3.
The two time intervals are selected to explore the short-term and long-term effects. 
The dependent variables are the total duration (in minutes) played in the target week, as well as the duration spent in social and progress sessions. 
Because the target variables are count data, all with a distribution for which the variance is higher than the mean, we use negative binomial regression. 
The descriptive statistics of the target variables are shown in Table \ref{tab:target_nw_3m_stats}.

\begin{table}[]
    \centering
    \begin{tabular}{lcccccc}
        \toprule
         & \multicolumn{3}{c}{\textbf{T2}} 
         & \multicolumn{3}{c}{\textbf{T3}}  \\
         \cmidrule(lr){2-4}\cmidrule(lr){5-7}
         & Total & Social & Progress& Total & Social & Progress \\
         \midrule
         Mean $\pm$ std & $351 \pm 617 $ & $190\pm 440 $ & $243 \pm 382$ & $266 \pm 575$ & $162 \pm 430$ & $178 \pm 348$\\
         Min -- Max & 0 -- 9,662 & 0 -- 8,972 & 0 -- 4,652 & 0 -- 9,669 & 0 -- 9,506 & 0 -- 5,290\\ 
         \bottomrule
    \end{tabular}
    \caption{Statistics of the duration (min) each player spends the following week (T2) and the week in three months (T3) from our target data period.}
    \label{tab:target_nw_3m_stats}
\end{table}

\begin{table}
    \begin{tabular}{lllllll}
    \toprule
    DV: Duration (min) & \multicolumn{2}{c}{Total} & \multicolumn{2}{c}{Social} & \multicolumn{2}{c}{Progress}\\
    \midrule
    Intercept
    	 & 3.047*** & \BarGreen{3.047} & 1.985*** & \BarGreen{1.985} & 2.650*** & \BarGreen{2.650}\\
    Group (friends) social
    	 & 8.635*** & \BarGreen{8.635} & 10.295*** & \BarGreen{10.295} & 7.830*** & \BarGreen{7.830}\\
    Group (friends) progress
    	 & 5.502*** & \BarGreen{5.502} & 5.119*** & \BarGreen{5.119} & 5.626*** & \BarGreen{5.626}\\
    Group (friends) social \& progress
    	 & 2.383*** & \BarGreen{2.383} & 3.476*** & \BarGreen{3.476} & 2.659*** & \BarGreen{2.659}\\
    Group (stranger) social
    	 & 2.707*** & \BarGreen{2.707} & 4.426*** & \BarGreen{4.426} & 1.890*** & \BarGreen{1.890}\\
    Group (stranger) progress
    	 & 2.611*** & \BarGreen{2.611} & 1.528*** & \BarGreen{1.528} & 2.828*** & \BarGreen{2.828}\\
    Group (stranger) social \& progress
    	 & -1.749*** & \BarRed{1.749} & -1.268*** & \BarRed{1.268} & -1.555*** & \BarRed{1.555}\\
    Solo progress
    	 & 0.401*** & \BarGreen{0.401} & -0.200*** & \BarRed{0.200} & 0.623*** & \BarGreen{0.623}\\
    \bottomrule
    \end{tabular}
    
    \caption{Engagement prediction results for the following week (T2--the second week of October 2020) using a Negative Binomial model. ***$p<0.001$}
    \label{tab:next_week_duration}
\end{table}
\begin{table}
    \begin{tabular}{lllllll}
    \toprule
    DV: Duration (min) & \multicolumn{2}{c}{Total} & \multicolumn{2}{c}{Social} & \multicolumn{2}{c}{Progress}\\
    \midrule
    Intercept
    	 & 3.305*** & \BarGreen{3.305} & 2.581*** & \BarGreen{2.581} & 3.305*** & \BarGreen{3.305}\\
    Group (friends) social
    	 & 7.935*** & \BarGreen{7.935} & 8.775*** & \BarGreen{8.775} & 7.935*** & \BarGreen{7.935}\\
    Group (friends) progress
    	 & 5.372*** & \BarGreen{5.372} & 4.929*** & \BarGreen{4.929} & 5.372*** & \BarGreen{5.372}\\
    Group (friends) social \& progress
    	 & 1.799*** & \BarGreen{1.799} & 2.413*** & \BarGreen{2.413} & 1.799*** & \BarGreen{1.799}\\
    Group (stranger) social
    	 & 2.007*** & \BarGreen{2.007} & 3.240*** & \BarGreen{3.240} & 2.007*** & \BarGreen{2.007}\\
    Group (stranger) progress
    	 & 1.937*** & \BarGreen{1.937} & 1.407*** & \BarGreen{1.407} & 1.937*** & \BarGreen{1.937}\\
    Group (stranger) social \& progress
    	 & -2.611*** & \BarRed{2.611} & -2.153*** & \BarRed{2.153} & -2.611*** & \BarRed{2.611}\\
    Solo progress
    	 & -0.336*** & \BarRed{0.336} & -0.545*** & \BarRed{0.545} & -0.336*** & \BarRed{0.336}\\
    \bottomrule
    \end{tabular}
  
    \caption{Engagement prediction results for the week in three months (T3--the first week of January 2021) using a Negative Binomial model. ***$p<0.001$}
    \label{tab:three_months_duration}
\end{table}

\paragraph{Findings.} Table \ref{tab:next_week_duration} and Table \ref{tab:three_months_duration} display the regression results in the short term---on the second week of October 2020 (T2)---and in the long term--- on the first week of January 2021 after three months (T3), respectively.
There is little difference between the short-term and long-term results. 
In comparing results between target variables, we see that prior social engagement is a stronger predictor of future social and progress engagement. 
This suggests that players who engage frequently in social and/or progress activities will continue to do so in the future.

Across all target variables and time periods, we observe that the duration spent in social activities spent with friends is the largest positive predictor of future player engagement, which is almost consistently followed by the duration spent in progress activities with friends. 
Social activities spent with strangers also increase future engagement but to a lesser degree. 
The coefficients of solo progress activities are minimally, but significantly, positive for future total engagement and future progress engagement predictions but negative for future social engagement in the short term. 
This could be explained by the existence of a small fraction of players who prefer to play the game solo. 
However, the coefficients of solo progress activities for engagement prediction in the long term (three months) are all negative, suggesting that solo players tend to not stay in the long run.
Nevertheless, the coefficients for social activities dominate future engagement prediction in all regards.
From this, we conclude that social activities, especially those with friends, are strong predictors of future engagement (\textbf{RQ1}).

\section{The Effects of Generosity}\label{sec:generosity}

To answer \textbf{RQ2} and \textbf{RQ3}, we examine whether acts of generosity have an impact on gameplay. 
We operationalize generous acts as candle giving, a defining characteristic of \textit{Sky} gameplay serving as the primary mechanism to facilitate friendship and social interactions among players.
Despite its importance in \textit{Sky} gameplay, we find that most players rarely experience candle exchanges---only 48\% of players received a candle in T1, and 60\% of the players who continued playing during T2 received a candle in T2.

\subsection{Preliminary Evidence of Contagious Giving}

In this analysis, our goal is to test whether we could find similar results to the studies described in \S\ref{sec:cont_gen} and specifically quantify the differences in generalized reciprocity versus third-party influence.
Quantifying this effect in a virtual world could help shed light on the expression of the phenomenon in the real world. 
We divide the players by their player interactions in their first session (direct or indirect) generosity and observe their generosity (\textbf{RQ2}) and engagement (\textbf{RQ3}) in the next session.
These divisions were:

\begin{enumerate}
    \item Players who themselves received a candle.
    \item Players who witnessed a nearby player give a candle.
    \item Players who witnessed a nearby player receive a candle.
\end{enumerate}

\begin{figure}
    \centering
    \includegraphics[width=0.95\linewidth]{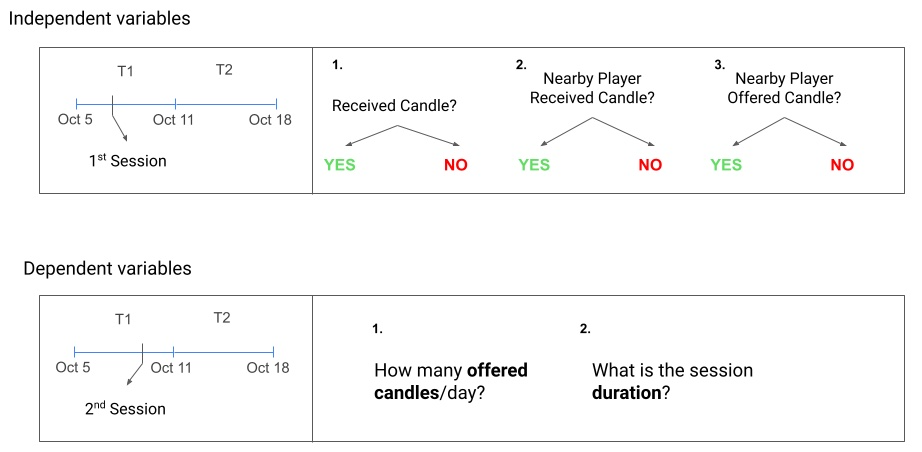}
    \caption{Experimental setup for modeling contagious giving.}
    \label{fig:cont_giv}
\end{figure}

The first one tests for generalized reciprocity and the next two tests for third-party influence.
Divisions (2) and (3) are very similar since in order to witness a candle received one has to witness a candle given. 
We measure both in case one of the involved nearby players was outside of the player's nearby threshold, thus not registering as a nearby player.
Overall, we expect the results for divisions (2) and (3) to be nearly identical. 
Our experimental setup is illustrated in Figure \ref{fig:cont_giv} and the results are shown in Table \ref{tab:social_results_candle}.

\begin{table}[]
    \centering
    \begin{tabular}{lllll}
        \toprule
          & \multicolumn{2}{c}{\textbf{Candles Offered/Day}} 
          & \multicolumn{2}{c}{\textbf{Session Duration }} \\
        \cmidrule(lr){2-3} \cmidrule(lr){4-5}
        {Event Occured} & \multicolumn{1}{c}{\cmark} &  \multicolumn{1}{c}{\xmark} & 
              \multicolumn{1}{c}{\cmark} &  \multicolumn{1}{c}{\xmark} \\
        If Player Received Candle          & 9.78*** & 7.63*** & 34.7*** & 25.8***  \\
        If Nearby Player Received Candle   & 7.99*** & 7.85*** & 38.5*** & 26.2***  \\
        If Nearby Player Offered Candle    & 8.01*** & 7.85*** & 38.3*** & 26.2*** \\
        \bottomrule
    \end{tabular}
    \caption{Results of subsequent session candle offering and duration after receiving or observing candle exchange. ***$p<0.001$. See Figure \ref{fig:cont_giv} for an outline of the experimental setup.}
    \label{tab:social_results_candle}
\end{table}

To test whether generosity is contagious, we analyze how receiving or observing candles influences how many candles players gave out in their subsequent sessions using two-sided Mann-Whitney $U$ tests (Table \ref{tab:social_results_candle}). 
We normalize the number of candles given by the length of the session because players who had longer sessions could have more opportunities to give candles. 
In this case, generalized reciprocity, i.e., the player receiving a candle personally, led to $28\%$ more candles given in their next sessions. We also see that even observing nearby candle exchanges can lead players to be significantly more generous in their next sessions.
Given that candle exchanging is very sparse---most players don't exchange candles in a session---this is a very meaningful result.
We did not take into account the seniority of players when doing this analysis, so it is possible that newer players do not know how to exchange candles until they were shown how to do it. 
In the following section \S\ref{sec:give_gp}, we address this limitation.

We observe very similar distinctions in the duration of playtime in the next session.
Players who received a candle in their first session played over two times longer in their subsequent session than those who did not receive a candle. 
Even observing a candle exchange led to increased playtime as well, although not of the same magnitude as a direct recipient. 

This section provides a meaningful initial answer to the impact of generosity: both receiving and observing generosity increases a player's own generosity in the future (\textbf{RQ2}) and engagement in the game (\textbf{RQ3}).

\begin{figure}
    \centering
    \includegraphics[width=0.85\linewidth]{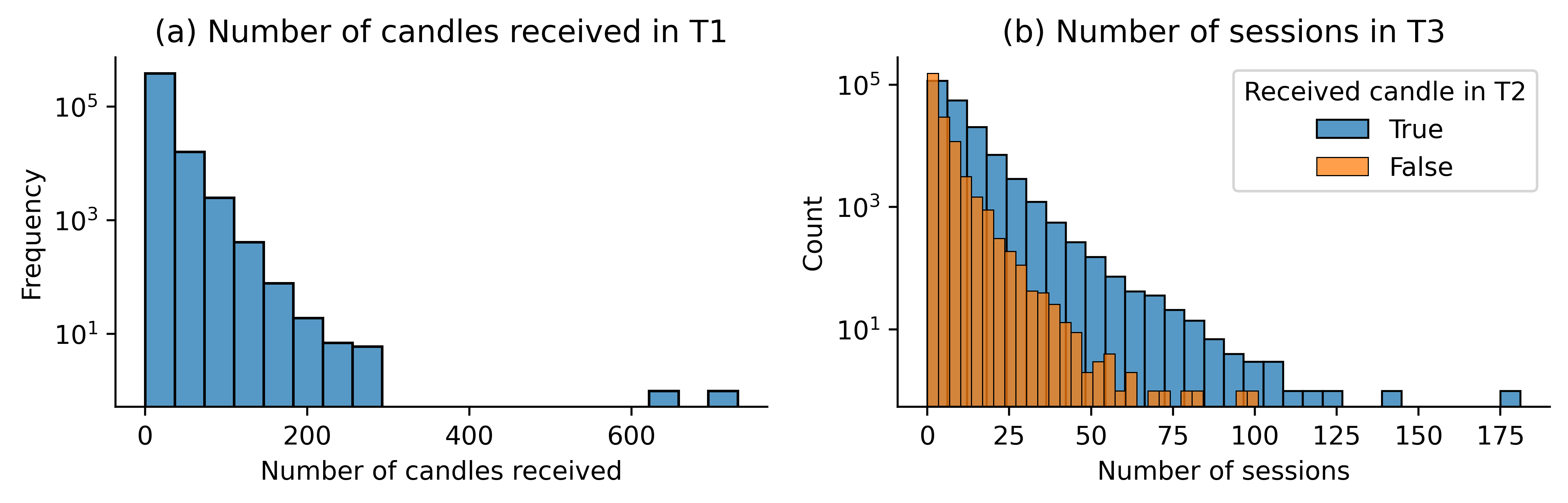}
    \caption{(a) Distribution of the number of candles received in the baseline period (T1). (b) Distribution of the number of sessions players had in the outcome period (T3), depending on whether they were given a candle in the treatment period (T2).}
    \label{fig:candle_retention_stats_w1-3}
\end{figure}

\subsection{Impact of Generosity on Gameplay} \label{sec:give_gp}
While the above analysis offers preliminary evidence that those who were exposed to generous acts are more likely to be generous and to engage more in the game in the future, this may be confounded with individual proclivities for social interactions and experience in the game, among other things. In order to more rigorously examine the impact of generosity in \textit{Sky}, we use a quasi-experimental approach called \textit{propensity score matching} \cite{imbens2015causal,rosenbaum1984reducing} to take into account individual characteristics for analyzing generalized reciprocity.

More specifically, we wish to compare the future outcomes $Y_i$ of player $i$ when $i$ was offered a candle (treatment $T=1$) versus when $i$ was not offered a candle (treatment $T=0$). 
However, there is only one real observation per player and the counterfactual outcome is unobservable.
We estimate the causal effect with propensity score matching, which enables the comparison of outcomes in distinct but comparable (\textit{matched}) groups of Treated and Control players. 
We consider the same three time periods of data described above in Figure \ref{fig:times}.
The first time period is the baseline period (T1), the second is the treatment period (T2), and the third is the outcome period (T3). Our goal is to compare players' outcomes in T3 based on their treatment in T2 by matching similar players together using their T1 covariates. 
The treatment is defined as whether the player received at least one candle from another player, and the outcomes include whether a player returned, how long they returned for (number of sessions played and total duration played), as well as whether they gave out candles in T3. 
After excluding players who did not play in T2, we are left with 405,000 players with 203,000 (50\%) Treated and 202,000 (50\%) Control during T2. 
Figure \ref{fig:candle_retention_stats_w1-3} illustrates the distribution of the number of candles they received in T1 and the number of sessions they played in T3 depending on their treatment status.

\paragraph{Propensity scores} 
To estimate the propensity of a player being offered a candle in T2, we fit a logistic regression model to predict whether a player will receive a treatment (i.e., be offered a candle) in T2 using their covariates in T1.
We describe below the four sets of covariates we use, which are aggregated per player:
\begin{itemize}
    \item \textit{Seniority}: the total number of days, duration, and sessions recorded for each player since the beginning of \textit{Sky}. These features are zero if they began playing in T1.
    \item \textit{Social engagement}: the number of social sessions and the total duration spent in social sessions.
    We also record the number of candles, gift messages, and chat messages sent or received, as well as the number of sessions in which hand-holding took place.
    \item \textit{Progress engagement}: the number of progress sessions and the total duration spent in progress sessions. 
    We also record the amount of wax farmed and the number of wing buffs that were collected or dropped.
    Finally, we record the number of quests completed.
    \item \textit{Nearby players}: the total number of nearby players and new friends made across all sessions. 
    We also calculate the mean seniority of nearby players and collect the number of candles received or offered nearby. 
    \item \textit{Play types}: the number of sessions spent in solo, group (friend), or group (stranger) play types. 
\end{itemize}

The covariates ensure that matched players share comparable characteristics of tenure, progression, sociability, and nearby players. 
We remove 36,000 outliers (8\% of total) whose propensity scores are more than two standard deviations away from the mean \cite{saha2019social}.

\paragraph{Matching} 
Using the propensity scores, we find groups (or pairs) of players that are most statistically similar in terms of their covariates. 
To avoid biases with exact matching \cite{olteanu2017distilling,king2019propensity}, we stratify the players into 100 strata of equal width. 
Strata with less than 100 of Treated or Control group of players are discarded to ensure we have enough data points per strata \cite{de2016discovering}, leaving us with 91 valid strata with 168,000 Treated and 199,000 Control players. 
To measure the quality of our matching, we apply two-way ANOVA on all covariates and find that all statistically significant covariates prior to matching are no longer significant \cite{rosenbaum1984reducing}. 
This shows that our stratified matching yields Treated and Control groups that are balanced in terms of covariates.

\begin{figure}
    \centering
    \includegraphics[width=\linewidth]{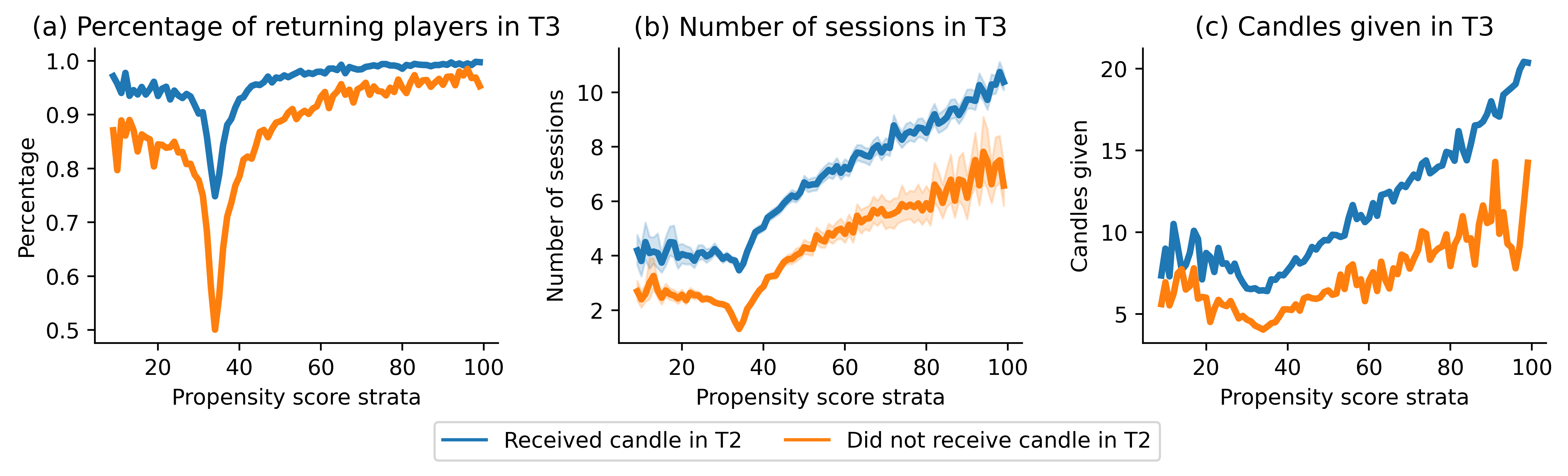}
    \caption{The outcome in T3 as a function of propensity score strata for the Treated and Control players.}
    \label{fig:pms_result_plots}
\end{figure}

\paragraph{Findings} 
We consider three measures of outcomes in T3: whether the player returned in T3, how many sessions they returned for, and how many candles they gave out. 
The first two outcomes measure their retention, while the last measures their inclination to be generous. 
We illustrate the outcome results as functions of the propensity score strata in Figure \ref{fig:pms_result_plots}. 
In Table \ref{tab:pms_results}, we show summary statistics of the Treated and Control players and compute the average treatment effect (ATE) and effect size (Cohen's $d$) across all population strata. 
We also perform paired $t$-test and $KS$-test on the differences in outcomes between the Treated and the Control players per strata. 

Our findings show that being treated with a candle in T2 significantly increased players' retention and their likelihood to be generous. 
In particular, 96\% of Treated players returned, compared to 88\% of Control. 
On average, being treated increased the number of sessions played in T3 by 2.27 sessions.
These findings reveal the drastic effect that generosity can have on gaming engagement. 
Importantly, receiving a candle in T2 more than quadruples the number of candles the players gave out in T3, further indicating that generosity is contagious. This section provides evidence to show that receiving generosity can significantly impact a player's capacity to be generous in the future (\textbf{RQ2}) and also that receiving generosity can lead to higher future gaming engagement (\textbf{RQ3}).
\begin{table}[]
    \centering
    \begin{tabular}{lcccccc}
        \toprule
         \textbf{Outcome} & \textbf{Treated} ($\mu$)  & \textbf{Control} ($\mu$) & \textbf{ATE} & \textbf{Cohen's d} & \textbf{t} & \textbf{KS}  \\
         \midrule
         Returned (T/F) &  0.96 & 0.88 & 0.08 & 0.11 & 14.81*** & 0.51*** \\
         Number of sessions & 6.67 & 4.40 & 2.27 & 1.02 & 37.39*** & 0.44***\\
         Number of candles given & 11.40 & 7.34 & 4.06 & 1.03 & 17.97*** & 0.45***\\
         \bottomrule
    \end{tabular}
    \caption{Average outcomes in T3 for Treated (received candle in T2) and Control (did not receive candle in T2) players as well as the ATE, Cohen's d, paired $t$-test, and KS test results. ***  $p<0.001$}
    \label{tab:pms_results}
\end{table}

\section{Discussion}\label{sec:discussion}

This paper examines the importance of social contagion and contagious generosity in the context of multiplayer online games, a relatively under-explored research area \cite{yee2006psychology,taylor2009play}. 
Using a large-scale dataset from a popular open-adventure social game, \textit{Sky}, we empirically analyze the multi-faceted ways in which social play can impact players' actions. 
We find that both directly giving and receiving acts of generosity---\textit{generalized reciprocity}---and observing generosity without reciprocation---\textit{third-party influence}---can impact players' generosity towards others in the future. 
Further, generosity has an impact on player retention and progression in the game, suggesting that generosity increases player satisfaction with the game.

The creator of \textit{Sky} states the game is one ``about compassion and generosity... It tries to evoke the bright side of humanity over the dark or grey in an online game'' \cite{apple2020behind}. That said, unlike many other MMORPGs, social interaction in \textit{Sky} is not strictly necessary.
A player can enjoy and progress in the game without having to interact with anyone.
In other words, the social play we observe in this game must purely be driven by the players' own desire to socialize. 
In many ways, relationships in \textit{Sky} function very much like those in the real world. The game entices players to engage socially by offering unique abilities and features that can only be unlocked through social play. 
Our results strongly suggest that social play, especially quality social play with friends, has an outsized impact on a player's future engagement in the game and desire to be generous. 
Our finding that social play and generous acts are contagious can have important implications for game design as well as other social settings.

\paragraph{Game playing is socially contagious.} 
Our work reveals that players exert significant influence on each other when they play together socially. 
This finding was replicated in the immediate term, the short-term, and the long-term.
Specifically, playing with others significantly increases a player's future engagement, both for social engagement and for progress engagement. Further, between playing with friends and playing with strangers, the former has a larger impact on future engagement. 
Playing solo is one of the most negative predictors of future engagement. 

\paragraph{Generosity is contagious.}
Candle exchange is the primary method to initiate friendship and unlock special social interactions in \textit{Sky}. 
Candles require time-intensive harvesting and also function as a virtual currency in the game.
However, the act of giving a candle is inherently an act of generosity because the favor does not have to be returned. 
This brings a unique opportunity for us to explore the effects of generosity. 
Our analyses reveal both generalized reciprocity and third-party influence exist \cite{voiskounsky2004playing} are readily observed in \textit{Sky}. 
That is, players are more likely to be generous if they either benefited from the generosity in the past or simply observed generous acts. 

\paragraph{Contagious generosity impacts future engagement.} 
Not only is generosity contagious, but it also impacts future engagement. 
Those who benefited from or observed generous acts spend more time playing the game in the future.
There are a number of explanations for this observation. 
First, as the generous act of candle giving serves as a gateway to special features and abilities in \textit{Sky}, players may want to play more because they enjoy the special friendship privileges they unlocked.
Second, players may be drawn to the quality of friendships they formed, which could either develop as a desire to play with their existing friends, or a desire to form new connections.
Third, studies have shown that people enjoy reciprocating generosity or paying it forward \cite{tsvetkova2014social,pressman2015pif}; therefore their continued gameplay could be a manifestation of their urge to extend the generosity. 
Though why humans remain generous and altruistic towards others remains largely unsolved, many believe that generosity leaves a \textit{feel-good} sensation in people \cite{sommerfeld2010subjective}. 

\subsection{Implications}
Our work supports several important implications, which we detail below.

\paragraph{Social play in game design.} Social play holds a spot of historic importance in game design. 
In 2007, \textit{Star Wars Galaxies} purposefully created virtual ``third places''---a designated place where people can voluntarily and informally gather for pleasure---to allow for organic social interactions to take place \cite{ducheneaut2007virtual}.
One key observation is that these spaces don't \textit{force} players to interact socially, they merely facilitate the interactions.
In fact, designated areas or levels of solitude in these ``third places'' actually give players a meaningful choice on how and when to socialize.

This philosophy could be encapsulated by ``player dynamics,'' a term coined by \citet{volldynamics2020}.
These are emergent interactions created by players themselves, or more broadly, actions outside of the control of game designers. 
Gift-giving and friendship formation are game mechanics that allow for emergent player dynamics in \textit{Sky}.
We show, empirically, that these interactions increase future gameplay. 
Thus, a natural way to boost engagement is to allow room for these generous player interactions to take place.
 
Our work suggests that adding elements of generosity to games will promote pro-social behaviors and increase player retention. 
We thus propose to supplement the ``player dynamics'' game design philosophy with generosity \cite{volldynamics2020}.
The process of how players perform generous acts should be considered from the beginning design phases of a game. 
In particular, generosity should feel genuine to the players and not contrived or forced.
For example, game designers could create NPCs (non-player characters) who act generously towards players or offer rewards to players who initiate generosity. 
However, this should be done carefully and tactfully to avoid players questioning the authenticity of generous behaviors and alienating the player base. 
To summarize, this work shows that generosity is an important aspect of social play in games and should be considered throughout the game design process.

\paragraph{Fostering virtual friendships and combating toxicity} 
In this work, we choose to focus on the positive side of online social interactions, foregoing an analysis or discussion of toxicity.
Toxicity in online mediums has long been a topic of research in social media \cite{salminen2018anatomy,bellmore2015five} and online games \cite{hilvert2020m,dowsett2019effect,adinolf2018toxic}. 
The shield of anonymity and the lack of barriers can enable toxic behavior online, negatively impact users' mental well-being.
Conversely, a benefit of open-ended social interaction, when executed correctly, is that it can amplify a players' engagement in the game.
We show that more often than not, players are positively impacted by social interactions. 
Our results confirm that online social relationships can expand social capital both at the collective level (enhancing generosity) and at the individual level (enhancing engagement).
Ultimately, there are desirable results to be reaped from social interactions that foster friendship, which we could also utilize to combat online toxic relationships.

\paragraph{Contagious generosity in other virtual worlds}

Although it appears that this may be a niche case for a specific fantasy game, there are many use cases of our findings in other games and social platforms.
It is not a stretch to see how these findings could be applied to multiplayer online games such as \textit{LoL} \cite{lol}, which employs a randomized loot system to increase revenue \cite{loot}.
In this system, players earn \textit{loot boxes} that have a certain chance of containing cosmetic upgrades that can normally only be obtained by spending real money.
We argue that combining loot and generosity opportunities can further increase player retention and satisfaction. For instance, if players have unwanted loot in their inventory, they could gift it to other players. This could potentially both propagate generosity via social contagion as well as increase player retention rates. 

Despite the arguments presented for transferability to other contexts presented above, one may still note \emph{Sky} does not \emph{require} verbal communication to form friendships and exchange gifts, which is different from many virtual worlds. 
\textit{Sky} players can communicate exclusively through candles or emotes; however, at the same time, text-based language communication does exist and plays a nontrivial role in player interactions.
Friends on \textit{Sky} can chat and communicate in the game after establishing a friendship through candle exchange. Additionally, there are virtual benches and tables that players can sit at to chat without being friends. Such mechanisms render communication in \textit{Sky} more similar to other contexts. 

Beyond games, generous behaviors also occur on social networks. For instance, \textit{Twitch} \cite{twitch} is a popular game streaming platform where individuals can follow and subscribe to streamers. 
Beyond supporting streamers through individual contributions, viewers are encouraged to gift subscriptions to other random viewers at the time the subscription is purchased \cite{gift-subs}. 
This becomes a win-win scenario in which the streamer is supported by additional viewers and the viewers experience generosity. 
\textit{Twitch} is a great example of a virtual social network where gifting and generosity are central pillars of community interaction. 
Virtual gift-giving on \textit{Facebook} has also been shown to result in both direct reciprocity and paying the gift forward \cite{Kizilcec2018social}.
In addition to these typical instances of gifting, one could also argue that interactions such as retweets, likes, follows, comments, etc. on \textit{Twitter}, \textit{Facebook}, and \textit{Instagram} are also displays of generosity. They are avenues in which people can relay endorsement and support to others. 
Such interactions need not be reciprocal but they often are, such as the case of the ``follow for follow`` strategy on \textit{Instagram} \cite{virtanen2017follow}, in which users are likely to reciprocate the following to the user who followed them first. 
The findings of this study thus carry many implications for understanding the contagion of generosity in other virtual worlds of interactions.

\paragraph{Implications for the real world}
Since 2020, our lives have increasingly pivoted from the physical world to the virtual world. 
Due to the COVID-19 pandemic, online interactions have become the sole medium for many people to maintain professional and personal connections. 
The line between what is virtual and what is not becomes increasingly blurred. 
The lack of face-to-face interaction leads to people feeling disconnected and isolated \cite{parent2021social,hwang2020loneliness}.
Many found it counter-intuitive or difficult to thrive in such an atmosphere \cite{Grant2021}. This perhaps is what gave rise to the popularity of many online multiplayer games as a way to maintain social connections \cite{forbesgames}.
One pertinent example was the explosion of \textit{Animal Crossing: New Horizons} since the COVID-19 pandemic \cite{Zhu2021}.
In this game, players craft a personal virtual world with opportunities for social encounters.
For example, players can trade scarce resources to create special items or send each other virtual gifts as tokens of appreciation. 
A study on \textit{Animal Crossing} playing during the pandemic found that visiting other players' islands is linked to less loneliness and anxiety \cite{Zhu2021}.
This example underscores the potential of utilizing virtual social encounters to improve the experiences and mental well-being of players in the physical world.

Ultimately, one could argue that the physical and the virtual worlds lie in one fluid continuum, both a part of the \textit{real world}.
Though many game researchers attempt to separate the virtual from the physical world, \citet{Lehdonvirta2010} questions the dichotomous approach in studying online games, asking, ``where does virtual space end and real world begin?'' 
With a plethora of forums, chat services, and video-sharing sites available to gamers, it is easier than ever for players to interact outside of the strict virtual space of the game.
While \textit{Sky} is a virtual fantasy game, its social aspects bear resemblance to mechanisms of friendships in real life, physical or virtual. 
Certain circumstances, such as a park bench or a picnic table, allow people to more easily share a common space and converse with strangers, much akin to the virtual chat benches and tables in \textit{Sky}.
As such, we are optimistic that our research regarding generous behaviors in online communities can translate into actionable insights for offline communities.

\subsection{Limitations and Future Work}
As our analysis was only done on one game, we did not empirically test the generalizability of these findings to other virtual worlds.
In particular, an important factor that we could not control is the demographic of people who play \textit{Sky}. 
It is possible that only the players who desire to socialize would be drawn to such a game, and less social players stopped playing before they could leave a meaningful signal. 

Another limitation of this work is that we used data collected only during the COVID-19 pandemic. 
During this time, online games and esports rose in popularity \cite{rise2020kim}.
This could potentially have led to a larger and less typical selection of players that were included in our study. 
That said, we believe that the conclusions we draw are still insightful and generalizable despite the unprecedented times this study transpired.

It is also possible that by only considering limited weeks of gameplay data, we are missing longitudinal trends or seasonal biases. 
Having only been released in mid-2019, \textit{Sky} is still a relatively nascent game with a growing player base. 
We aim to examine longer periods of gameplay data, such as a full year's worth of data, for future work.

Finally, we treat gaming sessions with friends as a blanket indication of quality playtime; however, it is relatively easy to form friendships in the game. 
Given the nature of candle exchanging, a friend session could be with an acquaintance---whom they just met in the game---or with a very good real-life friend they have known for a long time. 
Although challenging given the data resolution we have access to, a more nuanced approach would be to consider grading the quality of social playtime based on the established closeness between the two friends. 
It is also possible that playing with one or few friends has an entirely different social impact from playing with a large crowd of unfamiliar players. Thus, we propose examining the social impact of playing with varying numbers of friends in future work.

\begin{acks}
The authors are very grateful to Prof. Dmitri Williams and Steven Proudfoot (USC), for facilitating the connection with the TGC team, and to the rest of the game research group at USC Annenberg and USC Viterbi, including Prof. Tracy Fullerton, for their insightful feedback. 
The authors are indebted and grateful to Jenova Chen (TGC's President) for their enthusiastic collaboration, to 
Tina Lu (TGC) for her continuous support with data access and infrastructure engineering, and with the rest of the TGC team for their creative inputs and feedback on our work.
This work is partially supported by NSF (award no. 2035064) and DARPA (award no. HR001121C0169).
\end{acks}

\bibliographystyle{ACM-Reference-Format}
\bibliography{main}

\received{July 2021}
\received[revised]{November 2021}
\received[accepted]{April 2022}

\end{document}